\documentclass[sigconf]{acmart} 

\AtBeginDocument{%
  \providecommand\BibTeX{{%
    \normalfont B\kern-0.5em{\scshape i\kern-0.25em b}\kern-0.8em\TeX}}}

\setcopyright{rightsretained}
\copyrightyear{2025}
\acmYear{2025}
\acmConference[C\&C '25]{Creativity and Cognition}{June 23--25, 2025}{Virtual, United Kingdom}
\acmBooktitle{Creativity and Cognition (C\&C '25), June 23--25, 2025, Virtual, United Kingdom}
\acmDOI{10.1145/3698061.3726932}
\acmISBN{979-8-4007-1289-0/2025/06}

\usepackage{float}
\usepackage{booktabs}
\usepackage{graphicx}
\usepackage{placeins}
\begin{document}

\title{Human-Centered AI Communication in Co-Creativity: An Initial Framework and Insights}

\author{Jeba Rezwana}
\email{jrezwana@towson.edu}
\orcid{0000-0003-1824-249X}
\affiliation{%
  \institution{Towson University}
  \city{Towson}
  \country{USA}
}

\author{Corey Ford}
\email{c.ford@arts.ac.uk}
\orcid{0000-0002-6895-2441}
\affiliation{%
  \institution{University of the Arts London}
  \city{London}
  \country{United Kingdom}
}

\renewcommand{\shortauthors}{}

\begin{abstract}

Effective communication between AI and humans is essential for successful human-AI co-creation. However, many current co-creative AI systems lack effective communication, which limits their potential for collaboration. This paper presents the initial design of the Framework for AI Communication (FAICO) for co-creative AI, developed through a systematic review of 107 full-length papers. FAICO presents key aspects of AI communication and their impact on user experience, offering preliminary guidelines for designing human-centered AI communication. To improve the framework, we conducted a preliminary study with two focus groups involving skilled individuals in AI, HCI, and design. These sessions sought to understand participants' preferences for AI communication, gather their perceptions of the framework, collect feedback for refinement, and explore its use in co-creative domains like collaborative writing and design. Our findings reveal a preference for a human-AI feedback loop over linear communication and emphasize the importance of context in fostering mutual understanding. Based on these insights, we propose actionable strategies for applying FAICO in practice and future directions, marking the first step toward developing comprehensive guidelines for designing effective human-centered AI communication in co-creation.
\end{abstract}




\maketitle

\section{Introduction}
Human-AI co-creativity involves collaborative efforts between humans and AI in creative processes to produce artifacts, ideas, or performances \cite{davis2013human}, with research suggesting that the creativity resulting from this co-creation often surpasses the original intentions and creativity of the individual contributors \cite{liapis2014computational}. As generative AI (GenAI) systems with co-creative abilities, such as ChatGPT \cite{openaiChatGPTOptimizing}, DALL-E \cite{openaiDALLE}, Runway \cite{RunwayML}, and Midjourney \cite{midjourney}, become increasingly popular, human-AI co-creativity is gaining traction in mainstream applications. As humans closely interact with AI in co-creativity, interaction has been suggested as an essential aspect of effective co-creative AI in the literature \cite{bown2015player, kantosalo2020modalities}. As a result, the next frontier of co-creative AI should have good collaborative skills in addition to algorithmic competence \cite{rezwana2022designing}. However, designing effective co-creative AI presents several challenges due to the dynamic nature of human-AI interaction \cite{davis2016empirically, kantosalo2014isolation}. For example, co-creative AI must be capable of adapting to diverse and spontaneous interaction styles, allowing co-creative products to evolve dynamically across various stages of the creative process \cite{rezwana2022designing}. 

Communication is an essential component in human-AI co-creativity for the co-regulation of the collaborators \cite{bown2020speculative}. In this paper, we define \textit{\textbf{AI Communication}} as where an AI interacts with humans purposefully and directly to convey information. We distinguish \textit{AI Communication} as a component of the interaction between collaborators (humans and AI) rather than interaction with the shared creative product itself. This distinction aligns with the framework proposed by \citet{rezwana2022designing} in the co-creativity literature, where \textit{AI Communication} is separate from AI contribution types and creative processes that influence the final output. Our focus is specifically on direct communication, excluding indirect communication through sensemaking that occurs through creative contributions. For instance, in a co-creative drawing scenario, AI might provide feedback on users' contributions or proactively suggest design improvements, such as saying, ``The house you drew is a little out of proportion. Would you like to fix it, or should I adjust it for you?'' The user can choose to incorporate or disregard this suggestion, which distinguishes AI communication from AI's contribution to the creative process, such as drawing a tree next to the user's house. Our concept of AI communication aligns with the principles of explainable AI \cite{DARPA, gunning_2016}, especially in how AI communication can offer explanations within co-creative AI contexts \cite{xaixarts_workshop}.


To enable effective co-creation, \citet{mamykina2002collaborative} emphasizes that co-creative AI should offer feedback and critique on user contributions in a manner similar to human collaborators in a team setting. Research in HCI has shown that when technology directly communicates with people, it is often perceived as an independent `social actor' \cite{nass1994computers}, which enhances user perception of AI \cite{rezwana2022understanding}. As such, \emph{AI Communication} plays a crucial role in enabling co-creative AI to be viewed as an equal partner in the creative process \cite{mcmillan2002measures, rezwanacofi}, thereby improving user experience. Bidirectional exchange of ideas and information between users and AI can create moments of serendipity and novelty \cite{lawton2023tool}, which are essential components of creativity \cite{grace2015data, rezwana2024conceptual}. However, designing \emph{AI Communication} remains a complex challenge due to the dynamic and evolving nature of human-AI collaboration \cite{kantosalo2014isolation}, which involves managing nuanced interactions and accommodating diverse user needs \cite{weld2018intelligible}. This highlights a research gap for tools, such as frameworks, to facilitate effective \emph{AI Communication} in human-AI co-creativity.

Motivated by the research gap, in this paper, we introduce a Framework for AI Communication (FAICO) in co-creativity to improve user experience as the first step towards designing effective \emph{AI Communication} in co-creative contexts.
 Through a systematic review of interdisciplinary literature across HCI (107 full-length papers from the ACM Digital Library), we identified key aspects of AI Communication that we suggest should be considered in co-creativity and their influence on user experience, forming FAICO. We then present the findings of a preliminary user study in the form of two focus groups to gather feedback on FAICO in designing AI Communication for co-creative contexts. 
Our main contributions are:
\begin{itemize}
    \item A \textbf{Framework for AI Communication (FAICO)} for designing and interpreting \textit{AI communication}. FAICO identifies key aspects of \emph{AI Communication} that could be considered for effective co-creation and their influence on user experience.
    \item \textbf{Findings from a formative user study} with participants skilled in AI, HCI and design, providing feedback on the framework and insights on user preferences for AI Communication. 
    \item \textbf{Presentation of use-cases} of our framework into practical tools for the community and future research directions. 
\end{itemize}

The paper is organized as follows. We first give background on computational creativity research, human-AI interaction processes and communication in human-AI co-creativity. We then describe our literature review methodology and introduce our framework. Next, we present the user study in Section 5. We close by discussing our findings, suggesting improvements and presenting use cases of the framework.







\section{Background}

Co-creative systems originated from the concept of combining standalone generative AI (GenAI) systems with creativity support tools to create a system where both humans and AI take initiative in the creative process and interact as co-creators \cite{kantosalo2020role}. The terms human-AI co-creation and human-AI co-creativity are often used interchangeably in the literature, as both describe collaborative processes where humans and AI systems work together to produce creative outputs \cite{wu2021ai, wan2023felt, rezwana2022designing}. Mixed-initiative co-creativity (MICC) is a paradigm according to which creative agency cannot be ascribed to either the human or the AI alone but is instead a shared endeavor as both parties take the initiative \cite{dellermann2019hybrid, yannakakis2014mixed}. However, mixed-initiative creative systems are often used as a substitute term for co-creative systems in the literature \cite{yannakakis2014mixed}. Liapis et al. \cite{liapis2014computational} argued that when creativity emerges from human-computer interaction, it surpasses both contributors' original intentions as novel ideas arise in the process. In a creative setting, the role of interaction is crucial, as supported by Karimi et al., who found an association between human-AI interaction and different kinds of creativity that emerge from the co-creation \cite{karimi2020creative}. The design and evaluation of co-creative systems is still an open research question in the domain of co-creativity due to the spontaneous nature of the interaction between humans and AI \cite{davis2016empirically, kantosalo2020modalities}. To design co-creative systems in a way that considers the unique challenges of creativity research, Kantosalo et al. \cite{kantosalo2020modalities} emphasize that interaction design should be regarded as a foundational aspect. AI ability alone does not ensure a positive user experience with the AI \cite{louie2020novice}, especially where interaction is essential between humans and AI \cite{wegner1997interaction}. Furthermore, interaction has been proposed as a means to balance the agency between humans and AI in co-creativity \cite{moruzzi2024user}. However, research that focuses on the interaction between humans and AI in the field of co-creativity is still in its infancy \cite{rezwana2022designing}.



A few frameworks for interaction design for co-creative AI have been proposed in the literature \cite{Guzdial2019, kantosalo2020modalities,rezwanacofi}. Guzdial and Riedl \cite{Guzdial2019} describe human-AI interactions between humans, AI, and a shared artifact while focusing on turn-taking -- considering the actions both parties complete each time they contribute to the shared artifact. They suggest a distinction between turn actions (which contribute to an artifact) and non-turn actions (external aspects that might influence partners' contributing to an artifact). From this model, they suggest a distinction in the interaction style of co-creative systems in comparison to creativity support tools and generative systems: in co-creative systems, both partners must contribute information at every turn \cite{Guzdial2019}. \citet{kantosalo2020modalities} proposes a framework to facilitate analysis of co-creative systems by categorizing how human-computer collaboration occurs by focusing on three interaction levels: modalities (communication channels), styles (behaviors governing actions), and strategies (goals and plans). 
Later, inspired by previous Co-Creativity literature and computer-supported collaborative Work (CSCW) research, \citet{rezwana2022designing} presents the Co-Creative Framework for Interaction Design (COFI) for designing and evaluating interaction design in human-AI co-creativity. COFI presents the interaction design space by presenting both the aspects of interaction between co-creative AI and humans as well as the aspects of interaction with the shared artifacts. COFI highlights communication between humans and AI as a key interaction design component. More recently, Moruzzi and Margarido \cite{moruzzi2024user} proposed the User-Centered framework for Co-Creativity (UCCC) that provides a list of the key interaction dimensions in human-AI co-creativity that can be modulated by users to adjust their degree of control and to match their style. However, none of these frameworks specifically explores the dimensions of AI communication and their impact on user experience. In this paper, we extend the existing frameworks by focusing on the critical role of AI communication, identifying its key dimensions, and examining how these dimensions influence user experience.



\subsection{Communication in Human-AI Co-Creativity}
The exploration of communication with computer systems dates back to Alan Turing's proposal that a machine capable of communicating indistinguishably from humans could be considered intelligent \cite{french2000turing}. One significant challenge in human-AI collaboration, as highlighted in human interaction studies \cite{Clark1992}, is the establishment of common ground for communication between humans and machines \cite{dafoe2021cooperative}. We take the view that communication is an essential component in human-AI co-creativity for the co-regulation of the collaborators \cite{liang2019implicit, bown2020speculative}. In a human-AI collaboration, communication helps the AI agent make decisions in a creative process \cite{bown2020speculative}. For example, AI Communication in co-creation can improve user engagement, collaborative experience and user perception of a co-creative AI \cite{rezwana2022understanding, bryan2012identifying}. Co-creativity research suggests designing co-creative AI that communicate similarly to humans, acting as collaborators rather than as mere tools, foster a positive user experience \cite{biermann2022tool, rezwana2022understanding}.  

AI Communication is key in supporting many aspects of the user experience \cite{mcmillan2002measures}. For example, \citet{rezwana2022understanding} showed that incorporating AI Communication in co-creative AI improves user engagement, collaborative experience and user perception of a co-creative AI. AI Communication can also improve social presence and interpersonal trust \cite{bente2004social}. Bown et al. explored the role of dialogue between humans and AI in co-creation and argued that both linguistic and non-linguistic dialogues of concepts and artifacts are essential to maintain the quality of co-creation \cite{bown2020speculative}. Furthermore, research in HCI has also demonstrated that when technology directly communicates with humans, they are perceived as social entities \cite{nass1994computers}, highlighting the importance of AI communication. However, many existing co-creative AI systems include only human-to-AI communication and the AI can not effectively communicate with the human user \cite{rezwana2022designing}. Moreover, those co-creative AI systems that can communicate with humans have considerable challenges in achieving effective human-AI Communication \cite{shi2023understanding}.

Designing human-like communication has been described as an underestimated challenge of machine intelligence \cite{healey2021human}. HCI research shows that the way users talk in a human-AI conversation is similar to human-human conversation \cite{dev2020user}. While it is vital to use human communication and its ideas as a starting point for designing AI Communication, it should not impose permanent restrictions on AI Communication \cite{guzman2020artificial}. For instance, communication between artists is often based on intuition and non-verbal interaction \cite{healey2005inter}, rather than direct explanation-based discussions, which are more traditionally seen in AI for non-creative contexts \cite{xaixarts_workshop}. Identifying and designing the appropriate communication from AI is thus a key aspect of human-AI co-creativity research. Aspects of communication should also be chosen carefully because many elements of communication affect how humans interact with AI \cite{nass2005wired}, which may or may not always be appropriate depending on context. Furthermore, \citet{shi2023understanding} suggests that AI Communication should be more explainable to support collaboration. Among the benefits of receiving explanations from the AI, in addition to the mutual co-adaptation they facilitate \cite{van2021becoming} and the increased levels of trust in users it can promote \cite{schoeffer2022there, ahn2021will}, might be a higher level of user satisfaction with the output generated by the AI on its internal workings \cite{larsson2022towards}. In this paper, we introduce our first steps in creating a framework that could be helpful in guiding the design of AI communication, helping capture vital components to support effective interactions in human-AI co-creative processes.


\section{Framework Development Methodology: Systematic Literature Review}
In this section, we document the systematic literature review approach we followed to identify the key aspects of \textit{AI Communication} and their influence on user experience to develop our framework for AI communication.

\begin{figure*}[h]
    \centering
    \includegraphics[width=0.95\linewidth]{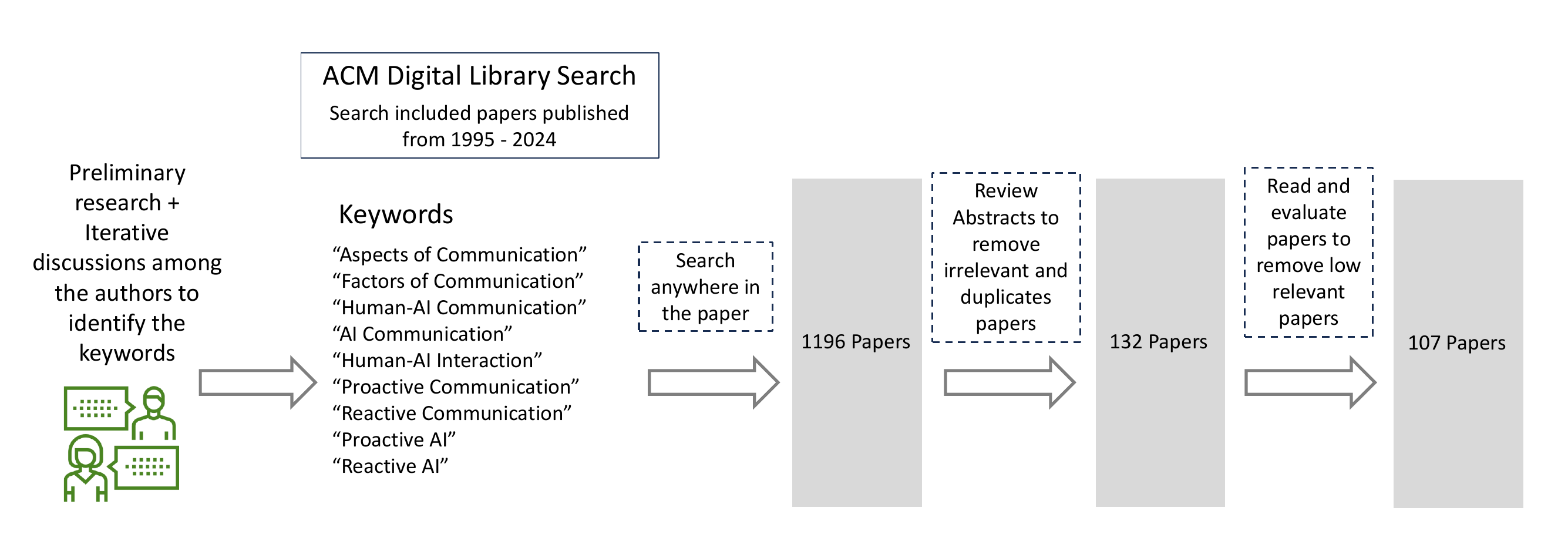}
    \caption{Flowchart showing the steps of the systematic literature review for developing the framework}
    \label{lit_review}
\end{figure*}

Figure \ref{lit_review} shows the methodology for our literature review. To identify essential aspects of \textit{AI Communication}, we used the ACM Digital Library (ACM DL) to connect to conferences where HCI, AI and creativity research is typically published and to build on reviews by others surveying creativity support \cite{frich2019} and human-AI co-creativity \cite{rezwana2022designing}. We considered documents published from 1995 until 2024 to consider recent articles. We narrowed our search to full-length papers (e.g. not tutorials or posters) to ensure we considered high-quality peer-reviewed studies. We also only considered papers in English that could be understood by both authors.

We used keywords aligned with our research goals to search for relevant academic publications. We began our SLR by identifying keywords through preliminary research, leveraging the research team's expertise. We chose the keywords following several weekly discussions where we reflected upon our own knowledge of existing research and our experiences of AI communication as co-creative AI researchers. Through iterative discussion, we selected nine keywords related to AI communication. The first two keywords related directly to communication: ``Aspects of Communication'' (561 results) and ``Factors of Communication'' (41 results). The next three are related to interaction and communication between AI and Humans: ``Human-AI Communication'' (47 results), ``AI Communication'' (117 results)  and ``Human-AI Interaction'' (279 results). The final four related to when the AI might communicate with humans: ``Proactive Communication'' (90 results), ``Reactive Communication'' (29 results), ``Proactive AI'' (19 results) and ``Reactive AI'' (13 results). 

After running the keyword-based search, we retrieved 1196 papers from the ACM DL. In the first screening round, the authors reviewed the abstracts to assess the relevance of the papers to AI communication and related aspects. To ensure consistency in the selection process, the authors reviewed the abstracts together to establish a consensus on inclusion criteria---the paper must directly or tangentially discuss different AI Communication aspects and strategies, and/or their influence on user experience. In the first round, from reviewing the abstracts of the papers and removing the duplicates, we had 132 papers that are relevant. In the second round, we conducted a full-text review and further filtered out papers with low relevance to aspects of AI communication and their influence on user experience, leading to the final corpus of \textbf{107 papers}.

We summarized the literature review on how co-creative AI should communicate using a shared spreadsheet with notes, insights and summaries. We then used affinity diagramming \cite{harboe2015affinity} to organize the summaries into related themes. The framework was developed in an iterative process of adding, merging, and removing key components of AI communication defined in the literature. Through iterative discussion, we created the first iteration of our framework for AI communication (FAICO). The framework developed and described below is primarily based on our literature review while also integrating relevant insights from \textbf{additional existing works} outside the scope of the review. The formation of FAICO follow a methodology similar to that used in other existing interaction frameworks for human-AI co-creation, such as COFI by \citet{rezwana2022designing} and UCCC by \citet{moruzzi2024user}. 





\section{Framework For AI Communication (FAICO)}

\begin{figure*}[h]
    \centering
    \includegraphics[width=0.9\linewidth]{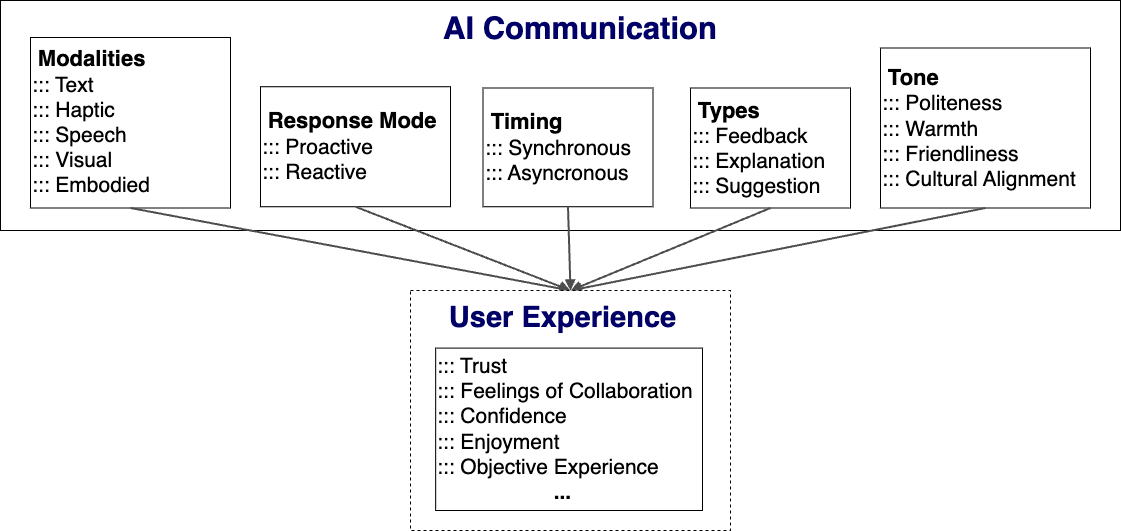}
    \caption{The Framework for AI Communication in Co-Creative Contexts}
    \label{framework}
\end{figure*}

In this section, we present the Framework for AI Communication (FAICO) (Figure \ref{framework}) in the context of human-AI co-creativity, introducing \textbf{critical aspects} to consider for AI communication and their influence on \textbf{user experience}. FAICO presents five key components of AI Communication that should be considered: \textit{modalities}, \textit{response mode}, \textit{timing}, \textit{types} and \textit{tone}. We discuss each aspect within FAICO below. We begin by defining each aspect and then describing its influence on user experience based on our literature review and additional relevant work. For each communication component, the first paragraph defines the component and cites the references that inspired its inclusion, while the second paragraph highlights studies that demonstrate its impact on various aspects of user experience.

\subsection{Modalities} 
According to FAICO, \textit{Modalities} refer to distinct channels through which AI agents communicate with humans, supported by HCI research \cite{karray2008human, rezwana2022designing}. \citet{rezwana2022designing} outlines the classification of modalities for AI-to-human communication in human-AI co-creation, encompassing text, speech, visuals, haptic, and embodied communication. In our framework, we adopt their classification of modalities for AI Communication.

\subsubsection{Connection to User Experience}
The literature demonstrates that communication modalities influence user experience. Text-based dialogue plays a key role in helping users perceive the AI's personality \cite{mairesse2007using} and enhances user satisfaction and enjoyment \cite{oh2018lead}. Speech-based communication improves perceived communication quality and trust, often performing better than visual cues in enhancing user experience \cite{zhang2024verbal}. For a real-time co-creative context people, however, people might need to rely on non-verbal cues, e.g, when musicians improvise together \cite{healey2005inter}. Nonetheless, verbal and non-verbal cues allow AI agents to convey their states effectively, which contributes to improved feelings of collaboration and fosters effective human-AI partnerships \cite{anzabi2023effect,bryan2012identifying, sonlu2021conversational}. Haptics and visual cues also play an important role, enhancing users’ understanding of emotional qualities and fostering a sense of connection \cite{wang_haptic_2023, price_touch_2022, rodrigues_emoji_2022, nanayakkarawasam_icon_2023}. For co-creative contexts, reflection on emotional qualities may be essential to one's artwork \cite{candy_creative_2019}. Embodied communication improves the feelings of collaboration with perceived presence compared to text-based communication \cite{lim2024artificial} - and there is a large body of work exploring embodied cognition in creative HCI research, e.g., \cite{embodied2016}. Furthermore, multimodal communication, such as combining visuals with verbal interactions, has been found to improve user experience \cite{samek2023cosy}. For instance, the integration of voice, embodied communication, dialogue, and facial expressions provides users with a more accurate perception of AI, improving feelings of collaboration and social presence \cite{sonlu2021conversational, munnukka_2022}.

The influence of modalities in AI Communication varies between user demographics such as gender \cite{huang2021women}, age \cite{razavi2022discourse}, neurodiversity \cite{samek2023cosy}, as well as the context in which multimodality is introduced \cite{esau2023foggy}. 
For example, \citet{vossen2009social} found a gender-specific response to embodied communication. To improve the user experience for a diverse population, \textbf{multimodal interactions} significantly enhance the performance of AI compared to unimodal interactions \cite{samek2023cosy, park2021toddler}. 

\subsection{Response Mode}
\textit{Response Mode} refers to the approach used by an AI agent to initiate or respond to communication during a co-creation. AI can be either \emph{proactive} or \emph{reactive} in terms of how it responds. Proactive AI systems ``do not wait for user input but instead actively initiate communication'' \cite{van2021proactive}. On the other hand, when an AI is reactive, it communicates with users when prompted, for example, when clicking a button \cite{meurisch2020exploring}.

\subsubsection{Connection to User Experience}
Response mode influences user experience in different ways. Proactive AI, while generally favored in social interactions over reactive AI \cite{meurisch2020exploring}, has been perceived as annoying in several studies \cite{van2021proactive}, particularly in contexts where it interrupts users' immersion in creative activities \cite{csikszentmihalyi_flow_1990} where reactive AI might be preferable. Reactive AI can improve task performance over time and improve perceived usefulness \cite{kim2021ai, berry2006can}. However, proactive AI can improve perceived trust \cite{zhang2023investigating, kraus2022kurt,jain2023co} and encourage users to be more critical of AI output, which we found connected with human-centered AI guidelines on responsible design \cite{ozmen2023sixchallenges}. Being critical of AI might also help to spark reflection which is key to the creative user experience \cite{coreysRiCEPaper}.

The demographics of users and contexts also affect the different types of timing that participants wanted. \citet{pang2013technology} found that while most mental health app users preferred reactive AI, researchers like doctoral students were more open to proactive AI. Similarly, \citet{meurisch2020exploring} demonstrated that factors such as age, country, gender, and personality traits impact user preferences, with individuals with higher openness favoring proactive AI, and those lower in extraversion preferred reactive communication. \citet{luria2022letters} further showed that parents' desired level of AI proactivity varied depending on the situation, emphasizing the importance of context.

\subsection{Timing} 
\textit{Timing} refers to \emph{when} an AI system communicates with the user, describing whether interactions occur simultaneously or at different times. Timing has been indicated as a crucial aspect of communication in human-AI collaboration \cite{zhou2024impact, fan2022human}. AI Communication can happen either in a \emph{synchronous} (whilst co-creating) or \emph{asynchronous} (outside of working times such as notifications) manner \cite{fan2022human}. 

\subsubsection{Connection to User Experience}
Timing affects the user experience in many different ways. Synchronous communication is beneficial when immediate feedback and real-time collaboration are necessary \cite{lim2017analysis}. It's particularly useful in improvisational co-creation, such as the studio stage of choreography creation, where AI can engage with users in real-time \cite{liu2024dancegen}. Situations requiring a spontaneous and immediate exchange of ideas and feedback from AI can enhance the co-creation process and successful task completion \cite{el2022sand}. Asynchronous communication can be useful when users need time for reflection and deeper consideration of AI suggestions \cite{striner2022co,berry2024reactive}. Creative activities that benefit from thoughtful consideration \cite{coreysRiCEPaper}, such as reviewing creative writing or art and producing reflection, can find asynchronous communication to be most useful \cite{xia2023review,liu2024dancegen}.

Timing influences different user demographics and creative stages differently. Experienced users often prefer asynchronous communication from AI \cite{wu2021ai}, likely because it allows them to maintain an uninterrupted creative flow. Different stages of the creative process may require different timing of AI communication \cite{wu2021ai}. For example, the ideation stage can benefit from synchronous communication, but the execution stage, where AI works on time-consuming tasks, can use asynchronous communication \cite{hwang2022too, lin2022creative}.

\subsection{Communication Type} 

Communication type refers to the type of AI Communication with which AI systems convey information to users based on distinct purposes. Research suggests that the type of communication is one of the major dimensions in AI-mediated communication \cite{guzman2020artificial, hancock2020ai}. According to FAICO, types of AI Communication include: \emph{explanation}, \emph{suggestion}, and \emph{feedback}. \emph{Explanation} can be defined as representations of underlying causes that led to a system’s output and which reflect decision-making processes. \emph{Feedback} refers to communication in the form of critique, analysis, or assessment of a particular contribution, idea, or artifact. 
\emph{Suggestion} refers to recommendations or alternatives to the user contributions during collaborative processes that align with the user's creative objectives. 


\subsubsection{Connection to User Experience}
Research shows that communication type influences various phases of co-creation. Example-based suggestions are effective in the early stages, fostering ideation, while rule-based suggestions are better suited to promoting user understanding \cite{yeh2022guide}. Divergent suggestions are particularly useful during the initial phases to encourage creativity, whereas convergent suggestions are more effective later, helping to refine the co-created product \cite{karimi2020creative}. AI explanations also play a crucial role in shaping user trust. Clear reasoning enhances trustworthiness \cite{mehrotra2024systematic}, and social transparency (such as making social contexts visible) improves both trust and decision-making \cite{ehsan2021expanding}. However, while providing reasoning increases trust, disclosing uncertainty can diminish it \cite{vossing2022designing}. Excessive trust may lead to overreliance on AI, reducing human accuracy \cite{nguyen2018believe}, a problem further amplified by human-like behaviors, such as hesitations in chatbot responses \cite{zhou2024beyond}. Moreover, users tend to value explanations of overall decision-making more than detailed explanations of individual actions, particularly in team settings \cite{zhang2021ideal}.

Users' responses to AI suggestions are shaped by demographics and task-related factors, including self-confidence, confidence and alignment with the suggestion, and pre-existing beliefs about human versus AI performance \cite{vodrahalli2022humans}. Tailoring AI communication to align with users' cultural backgrounds further enhances interaction effectiveness \cite{zhao2024tailoring}.

\subsection{Tone} 
\textit{Tone} refers to qualities of AI Communication that determine whether the expression of the communication conveys a positive emotional intent or not. Tone has been suggested as an aspect that can affect human-AI team performance \cite{mallick2024pursuit, leong2023exploratory}. In our literature search, we found four influences on the perceived tone of AI Communication - \emph{politeness} (use of considerate, respectful language and behavior following social norms) \cite{colley2021}, \emph{warmth} (being approachable and affable in interactions) \cite{khadpe2020conceptual}, \emph{friendliness} (ability to foster congenial relationships) \cite{sun2024towards} and \emph{cultural alignment} (following practices and norms matching the expectations of those involved) \cite{tao2024cultural}. 

\subsubsection{Connection to User Experience}
The tone of AI communication significantly impacts user experience. \citet{colley2021} found that when virtual AI avatars in an autonomous vehicle simulation communicated acknowledgment of a user’s polite gestures (e.g., waving thank you), the positive perception of the AI increased, demonstrating how respectful behaviors and adherence to cultural norms enhance user experience. Warmth and friendliness also improve user satisfaction and perceptions of AI capability. For instance, \citet{gilad2021warmth} showed that increasing an AI’s warmth positively influenced user satisfaction, whereas \citet{poeller2023suspecting} found that overly positive messages were sometimes met with skepticism in video game contexts, emphasizing the importance of context in determining the appropriateness of tone. Similarly, AI that responds in a human-like manner with supportive and friendly communication fosters a more engaging and satisfying user experience compared to interactions that lack these qualities \cite{d2013autotutor}. For a co-creative context, the friendliness of tone would need to be considered to be conducive to creativity -- might an AI that is too friendly not add necessary pressure to support extrinsic motivation \cite{amabile_creativity_1996} and potentially help with creative blockers \cite{lewis_aixartist_2023}.

\citet{mascarenhas2009cultural} demonstrated that variations in AI agents' rituals based on cultural behaviors had noticeable effects on users. Likewise, users preferred conversational AI aligned with their cultural norms and were adept at identifying cultural differences in dialogue \cite{endrass2009culture}. Aligning the tone of AI communication with demographic factors such as age, gender, and cultural background can lead to more effective and satisfying interactions \cite{sicilia2024humbel}.

\section{Preliminary Formative Study}
To identify users' preferences for AI communication and gather feedback on FAICO, we conducted a preliminary study with two focus groups. These groups were designed to be exploratory rather than strictly evaluative, aiming to collect qualitative feedback and suggestions for improving the initial design of our framework. We emphasize the study as preliminary and do not claim that our findings will be universal for all users and user preferences. We focused on exploring AI communication within the contexts of co-creative writing and design to assess FAICO's applicability across different domains. The study was approved by the ethics review committees of both authors' universities.

\begin{figure*}[h]
    \centering
    \includegraphics[width=0.95\linewidth]{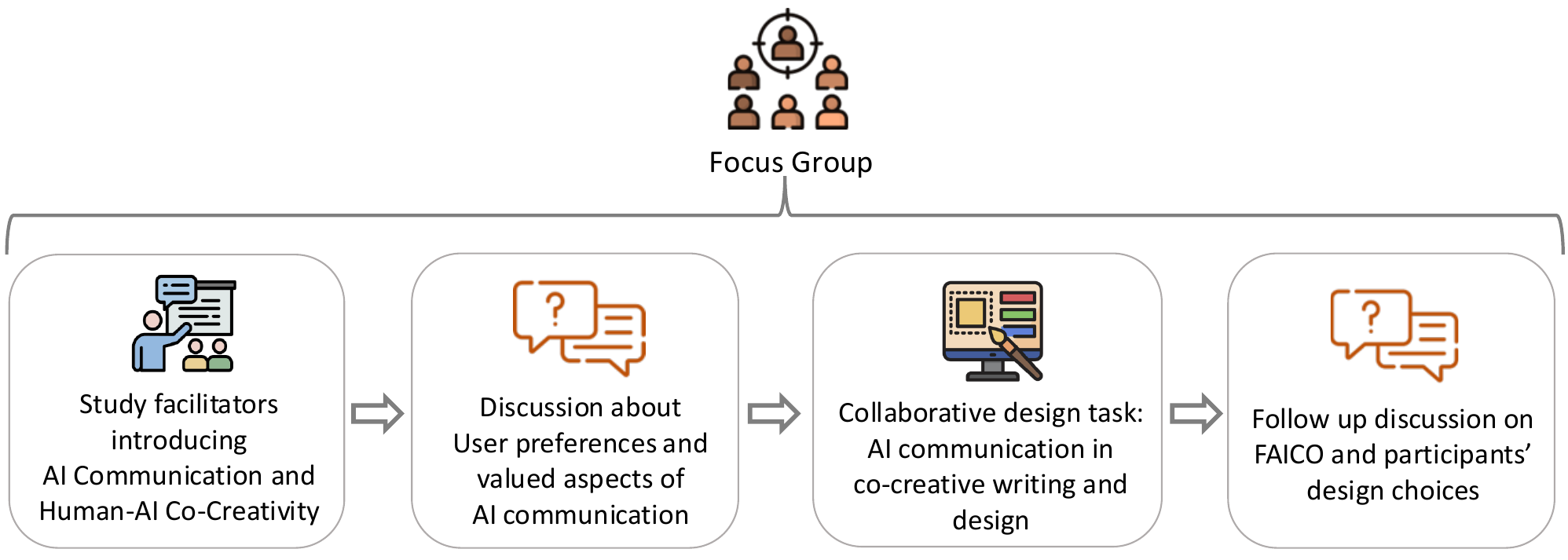}
    \caption{Study Procedure}
    \label{procedure}
\end{figure*}

\subsection{Participants}


We recruited 13 participants, with 7 in the first focus group (Table 1 in the appendix) and 6 in the second (Table 2 in the appendix). However, due to technical issues encountered by Participant 5 during the remote session, our findings are based on data from the remaining 12 participants. Participants' ages ranged from 22 to 39 (M = 29.9). Participants were selected through our personal and professional research networks based on their expertise. Eligibility criteria required them to be adults (18+) with expertise in AI, HCI, or Design. Given that human-AI co-creativity is inherently interdisciplinary \cite{yang2020reexamining}, we aimed to foster discussions across diverse perspectives by ensuring a balanced mix of skills in each focus group. We assessed skill sets through a demographic questionnaire (Appendix). Our participant pool included 4 individuals skilled in AI, 5 in HCI, and 3 in Design. Each participant had a minimum of three years of experience in academia, research, or industry within their respective fields. Participants were compensated with a \$20 Amazon voucher.


\subsection{Procedure}
The focus group studies lasted approximately 1 hour 30 minutes and were conducted as follows (summarized in Fig \ref{procedure}):
\begin{itemize}
    \item First, participants joined the focus group on Zoom\footnote{\url{https://zoom.us/}}, where their usernames were anonymized using identifiers such as P1, P2, and so on. We conducted the focus groups online to enable participation from individuals across different time zones and locations, ensuring greater convenience and inclusivity.
    \item An introductory presentation was then given by the researchers describing: (1) our definition of Human-AI Co-Creativity, (2) our definition of AI Communication, (3) examples of co-creative AI in collaborative writing, such as ChatGPT \cite{openaiChatGPTOptimizing}, and (4) examples of co-creative AI in design, such as DALL-E \cite{openaiDALLE} to ensure participants had a shared understanding of the case study areas and have common examples for discussion. A script was followed to maintain consistency between sessions.
    \item Next, to initiate discussion, the following opening questions were asked: (1) How do you want an AI partner to communicate with you in a co-creation?; (2) What aspects of AI Communication do you find valuable in the context of co-creativity, such as in co-creative writing and design?. These contexts were chosen to align with recent trends in co-creative AI and mainstream use of popular AI tools e.g. \cite{openaiChatGPTOptimizing,openaiDALLE,midjourney,RunwayML}.
    \item Next, participants were invited to join an online whiteboard \footnote{\url{https://webwhiteboard.com/}} for a collaborative design session. The whiteboard displayed an image of FAICO, providing a reference to support participants in the design process. 
    However, we did not explicitly ask participants to consider all of FAICO's components to investigate whether some components were discussed or considered more than others. Participants were then given two design tasks: (1) to design AI Communication in the context of collaborative writing and (2) in the context of collaborative design. Eight minutes were allocated for each design task, where participants contributed while discussing among themselves. 
    \item After the design tasks, participants were asked closing questions about (1) AI Communication aspects within FAICO that they considered in their tasks, (2) whether our framework influenced their perspectives on AI Communication in co-creation, and (3) broader feedback on our framework, e.g. how useful it was when designing AI Communication. 
\end{itemize}

\subsection{Data Collection \& Analysis}
We collected audio recordings from our focus group and the whiteboard designs, including sketches, notes and scribbles. The focus group discussions were transcribed using the transcription tool Sonix \cite{sonixAutomaticallyConvert}, which we then used for a Thematic Analysis. We followed Braun and Clarke's \cite{braun2012thematic} six-phase structure, moving from generating codes from the data to organizing these into themes. Both of the paper authors conducted coding independently and met later to agree on the final 39 codes in which to construct the primary themes. Note that we do not include additional quantitative details on how many participants supported each theme or inter-rater reliability metrics as our Thematic Analysis followed a reflexive approach \cite{braun_reflecting_2019} -- we see quantitative additions as undermining the interpretative nature of this approach, as supported by existing HCI scholarship \cite{braun_one_2021,soden2024,Crabtree_2025}.


\subsection{Findings}
From our thematic analysis, we generated six themes offering preliminary insights into our framework. We describe the themes below, illustrating our findings with quotes from the participants. 

\subsubsection{\textbf{Theme 1: ``The framework is helpful'' - General Perceptions of FAICO}}

\noindent This theme discusses participants' general, overall perceptions of FAICO for designing AI Communication. Participants discussed the utility of the framework as P10 underscored the practicality of the framework, noting, ``The framework is actually pretty helpful in thinking about how we want to design such tools.'' P8 echoed this sentiment, emphasizing the organizational value of the framework, stating, ``The framework makes it organized, focusing on including helpful components in the user interface.'' The analogy of the framework as a checklist resonated with participants, with P11 expressing, ``I'll definitely use it as a checklist... it's helpful.” P9 concurred, stating, ``It would be a good checklist.”

Participants acknowledged the impact of FAICO on their design considerations, giving varying levels of attention to different aspects of our framework. When asked about the components in the framework that were considered in their designs, most mentioned \textit{modalities} and \textit{types}. For example, 
P10 said, ``I looked into the modalities, suggestion type and a little bit of timing.” 
 Participants also discussed components they would not have considered unless they had the framework. For example, many mentioned that they would not have thought about AI \textit{tone} without FAICO; P6 said, ``Tone is very important... but it did not spring to mind until I saw it in the framework. I feel like that's something a lot of AI systems don't really consider, but I think it really changes someone's confidence, whether the AI is coming across warm or with good intentions rather than blunt.'' 

\subsubsection{\textbf{Theme 2: ``AI should use clear and simple language'' - User Preferences for AI Communication}}

\noindent This theme discusses users' expectations regarding how AI should communicate in creative contexts. 
Specifically, discussions arose regarding the details of explanations in a co-creative setting, as participants suggested that complicated explanations could make the user experience worse. P1 pondered, ``How much is the sensory overload?'' as ``trying to explain all AI decisions and AI generations might end up being a lot of new data for the user.'' 
On the other hand, participants expressed a desire for simple, short, and informative interactions to enhance co-creation without causing disruption. Participants also emphasized the importance of clarity in AI Communication. 
P12 stated, ``AI should use clear and simple language to convey the information, not unnecessarily very technical or very complex terms.''

Discussions also centered around the ethical considerations of co-creative AI in the context of AI communication, including concerns about transparency, trustworthiness, data collection, and empathy. For instance, P1 underscored the role of trust and transparency between humans and AI in co-creation, stating, ``The human-AI conversation needs to have a lot of trust because I allow you to understand me, and AI should allow me to understand it.”  Participants emphasized transparency in how AI utilizes the information it collects from users. P13 stated, ``AI Communication should involve transparency and users should know how it's using all the information.” 
P6 described how an AI lacking empathy in its communication might be perceived as disingenuous, saying:``It takes a lot to make AI bots, but people won't use it if they don't feel like it's considering their feelings as it communicates.”



\subsubsection{\textbf{Theme 3: ``It should be a feedback loop'' - Users want Human-AI Communication Loops}}

\noindent This theme focuses on user preference for human-AI communication loops in the form of dialogues over a linear, one-way communication approach. Participants criticized AI systems that generate outputs based solely on users’ initial prompts without seeking clarification. 
Participants advocated for a continuous human-AI communication loop, suggesting that meaningful collaboration is best achieved through ongoing interactions rather than a single prompt. In this context, P7 said, ``It should be a chat kind of thing, not just one single prompt.'' In the same context, P13 said, ``AI Communication should involve multiple rounds, multiple interactions, not just one. It should give you feedback for every step.'' 

Participants proposed strategies to make prompt engineering more guided using AI Communication. In this context, P4 suggested, "AI could give suggestions to educate the user on what type of prompt they to achieve their goals." P11 emphasized AI communicating ways to write effective prompts for non-expert users by saying, ``Help me learn how to write the prompt or communicate better because I'm not an artist.'' 
Participants discussed their preference for co-creative AI with conversing interaction, which they perceived as more partner-like, over one-shot interaction. P13 pointed out this contrast by saying, ``In DALL-E, you don't really feel like you are talking in a continuous manner with the AI. You're giving it some words and it's giving you feedback. Whereas in ChatGPT, you're continuing to have this conversation that's built on previous steps, and it responds as if it is a partner.” 
 Participants discussed the role of communication in human-AI co-creation in fostering mutual understanding, e.g. 
 P4 said: ``The communication is mainly about understanding each other... If we could improve understanding each other, the collaboration would be much easier.”



\subsubsection{\textbf{Theme 4: ``Communication should be according to the situation” - Context-based AI Communication}}

\noindent The theme revolves around the nuanced aspects of context-based AI Communication. 
Participants emphasized the importance of tailoring AI Communication to specific situations, interactions, and contexts. P12 articulated this, stating, ``It should be according to the situation, according to the interaction, previous interaction, according to the context.” 
P12 expressed the need for customizable options based on different communication contexts, such as academic language, casual conversation, or professional office email.
 P10 emphasized the role of urgency in AI Communication, stating, ``One very important aspect is how urgent a certain interaction is to me. If it is super urgent, the AI output needs to be exactly to the point. But if I have some time and I really want some ideas, abstract interactions from AI work as well.” P12 expanded on this, noting that the concreteness or abstractness of AI feedback depends on the creative phase. 
 
 The participants underscored the importance of timing and context in shaping divergent versus convergent AI Communication. P11 said, “I want the AI to give me more open-ended and creative responses at the beginning. But towards the end, when I'm refining the writing or the art, I just want it to follow my direction.” 
Cultural alignment in AI Communication also emerged as a crucial aspect, particularly after reviewing FAICO. P2 suggested, ``AI could ask - do you want to tone your paragraph or writing aligned with any specific culture?... The GPT model’s training data was kind of biased for Western culture.” Overall, the participants underscored the need for AI systems to be contextually aware, flexible, and culturally sensitive to enhance collaboration in creative endeavors.


\subsubsection{\textbf{Theme 5: ``It should have different options” - Users Want Flexibility in Configuring AI Communication}}

\noindent This theme revolves around participants' being able to configure and customize AI communication aligned with their preferences and creative goals. Participants articulated the importance of tailoring AI Communication to individual preferences. 
Specifically, participants exhibited diverse preferences on modalities, asking for flexibility and options for choosing the modality for AI communication. For example, P12 highlighted user flexibility by saying, ``There should be an option to choose how you want the AI to communicate. If you want to hear it, you can hear it. Or if you want to see it, you can see it.”  
Participants also critiqued existing co-creative AI systems, such as ChatGPT, for the lack of customization and flexibility. P7 noted, "ChatGPT has a specific pattern of interaction. And, I think it would be better if it would be more customized." 


Participants also wanted more precise control and flexibility over which parts of the shared product they wanted AI suggestions or explanations on. P11 said, "I want more flexible ways of pointing to where I want feedback. I want to highlight part of the text and tell the AI to do so." In the same context, P8 suggested having flexible options for the users as ways to control how AI should communicate with them in co-creation. Moreover, participants sought the flexibility to request AI responses to specific aspects of shared artifacts without discarding the entire output.



\subsubsection{\textbf{Theme 6: ``It will be able to connect with you” - Partner-like AI Communication Fostering Emotional Connection}}

\noindent This theme explores participants' consideration of affective aspects in AI Communication and the preference that the AI should communicate as a collaborator rather than a tool. Whilst advocating for a nuanced understanding of user preferences, some participants recommended that AI use user emotions to communicate in a better way. P6 stated, ``AI need to track human emotions to communicate in a much better way.'' Some participants considered the efficiency of collecting user preferences based on facial expressions rather than alternative modalities such as visual cues. 

To ensure effective collaboration, participants stressed the importance of human-AI communication mirroring human-human communication. P8 expressed the necessity for users to feel like they are collaborating with a partner when interacting with co-creative AI. 
Participants envisioned the role of AI as akin to a friend, where AI possesses a deep understanding of users on a personal level. 
P6 said, ``With your closest friend, you can just talk with your eyes and not have to say anything. So it's similar with AI in the sense that if it knows what sort of person you are, the type of creativity you want, and your personality type, it will be able to connect with you on a level beyond words.” P3 emphasized the need for equal collaboration between humans and AI, while P1 echoed this sentiment, asserting that treating AI as a mere tool is insufficient in a collaborative setting by stating, ``I like the idea of seeing AI as a collaborator rather than using it as a tool.''

\section{Discussion}

Despite the rapid advancement of generative AI in creative domains, many co-creative AI systems \textbf{lack effective AI communication} \cite{rezwana2022designing}, limiting the user experience \cite{rezwana2022understanding}. Designing effective human-AI interactions requires understanding the unique complexities of AI design and generating insights to guide future research \cite{yang2020re}. In response, this paper presents a novel framework for effective AI Communication (FAICO) which could be applied to improve user experience in human-AI co-creation. Through a systematic literature review, we identified key aspects of AI Communication and how they influence the user experience. We then presented a preliminary study with participants skilled in design, HCI and AI to understand their perceptions and gather feedback on FAICO. Below, we reflect on our framework based on the findings of the user study, suggest use cases for FAICO, and outline the limitations and future research directions.

\subsection{Reflections on FAICO}


Participants highlighted FAICO as a valuable tool for designing AI communication in co-creative contexts (Theme 1). They noted that FAICO helped them consider aspects they might not have otherwise thought about and suggested that it could serve as a checklist of key elements for designing AI communication to enhance the user experience of co-creative AI. Existing research shows that checklists are often used to guide design processes and act as a valuable design tool \cite{Kulp2017Exploring, Camargo2018Visual, Ji2006Usability}. These findings from our preliminary study support FAICO’s primary objective of guiding the design of effective human-centered AI communication.

In the focus groups, feedback loops were identified as a crucial aspect of AI communication (Theme 3), which aligns with research on human-human communication, where turn-taking fosters mutual understanding between participants \cite{Clark1992, healey2021human}. For example, participants highlighted the value of grounding mechanisms, such as AI helping in designing prompts through an iterative communication loop, improving mutual understanding or providing additional explanations through dialogue. This finding mirrors co-creativity literature, which underscores the importance of turn-taking and incorporating AI feedback in the creative process \cite{kellas2014rating2, Guzdial2019, rezwana2022designing}. Recent studies suggest that leveraging large language models (LLMs) could enhance nuanced and spontaneous human-AI communication, thereby improving collaboration \cite{miehling2024language, guan2023efficient, wan2023felt}. \textit{Future research} could explore strategies for recognizing user needs through communication loops, potentially utilizing LLMs, to foster a shared understanding between humans and AI. 

To inform the pattern and structure of AI explanations, participants expressed concern about sensory overload (Theme 2), with most favoring short, clear, and simple explanations. They also emphasized the importance of transparency in AI communication to build trust. However, our findings indicate that the level of AI explanation and ambiguity should vary depending on the phase of the creative process and the co-creation context—for instance, providing minimal explanations for time-sensitive tasks while allowing more detailed explanations for tasks that involve reflection and ideation (Theme 4). This finding resonates with the findings from \citet{lin2023beyond} that adapting AI interaction to different stages of the co-creative process can improve user experience. Research indicates that AI suggestions and explanations can help expand idea generation in the early creative stages, whereas more focused explanations become more effective in later refinement stages \cite{shaer2024ai}. Based on this theme, we expanded FAICO by adding \textit{Explanation details} as an essential component of AI communication \citet{rezwana2025improving} in the latest version of the framework. \textit{Future research} could explore how AI communication should change based on the creative phases and contexts.

The study revealed varying preferences for AI communication, emphasizing the need for flexibility. Participants strongly favored having multiple options for AI Communication modalities (Theme 5), allowing them to tailor their experience based on their preferences and goals. This aligns with research showing that diverse modalities enhance user engagement and collaborative experiences \cite{sonlu2021conversational, mairesse2007using, oh2018lead, rezwana2022understanding}. Additionally, studies suggest that users who customize their experiences through design toolkits perceive greater value in their creations compared to standard offerings \cite{franke2004value}. Beyond modality preferences, we observed differences in how participants with varying expertise preferred different types of AI communication. Participants with backgrounds in AI and HCI, but not necessarily in design, were less comfortable with ambiguous feedback. For example, P9, with expertise in design, preferred AI communication that incorporated some uncertainty to spark inspiration, whereas P11, with expertise in HCI, favored clear and direct communication. These findings might suggest that AI communication should be adaptable to users' expertise levels, balancing clarity and ambiguity to support creative divergence. \textit{Future research} should explore strategies for designing AI communication that effectively accommodates diverse expertise while maintaining flexibility across creative contexts.


Participants emphasized the importance of affective AI communication in making AI feel like an equal partner rather than just a tool in co-creative settings (Theme 6). This aligns with \citet{rezwana2022understanding}, who found that AI communication can enhance the perception of AI as a collaborator, leading to greater engagement and a stronger sense of collaboration. Drawing parallels between human-AI and human-human collaboration, participants frequently anthropomorphized the AI, reinforcing the idea that establishing a partnership dynamic is crucial in co-creation. This observation is supported by prior research, which suggests that perceiving AI as a collaborator rather than a tool enhances the overall collaborative experience \cite{rezwana2022understanding, biermann2022tool}. The emphasis on affective computing also resonates with studies advocating for empathetic AI partners to improve user experiences \cite{mutlu2009nonverbal, abdellahi2020arny}. \textit{Future research} should explore how to design AI communication that is affective and fosters a deeper sense of partnership in co-creative processes.


\subsection{Use-Cases For FAICO}
FAICO can be used to explore potential design spaces for AI communication in co-creative contexts. It also has applications in AI and HCI research, where it can guide the design of AI communication to improve user experience across various domains. Based on the insights discussed above, we outline three use cases for how FAICO can be applied by users, designers, and AI practitioners in future work.

\textbf{Design Cards:} We suggest that FAICO could be transformed into design cards to be used by designers and AI practitioners to better design AI communication to improve user experience. Design cards have long been recognized as valuable tools in HCI and design work, known to facilitate ideation, prototyping, and implementation \cite{lucero2016designing, hsieh2023cards}. Each card in the proposed set of five would focus on a specific component of FAICO, explaining its significance in AI communication and its potential impact on user experience. These cards would prompt designers with various possibilities for AI communication, providing them with valuable guidance that can be applied directly to their design context.

\textbf{Configuration Tool:} FAICO could be translated into a configuration tool within co-creative systems for users, giving them the autonomy to adapt and configure the communication to their preferences. For example, users could select from various modalities or response types—proactive or reactive—listed by FAICO. As our findings show that participants want flexibility over how they want their AI to communicate with them and their expertise shapes AI communication preferences, this could help to support AI systems' perceived value and engagement without prescribing a preference for a certain user type. Research shows that allowing users to experience different interface configurations results in interfaces that enhance user engagement and satisfaction \cite{hui2008toward}.

\textbf{Evaluation Tool:} FAICO can also be used as an evaluation tool to interpret and assess the design of AI communication in existing co-creative AI systems. Using FAICO, researchers could identify trends, strengths, and gaps in communication design, providing valuable insights into how existing systems approach communication and highlighting areas for improvement to enhance user experience and future research directions. For instance, FAICO can reveal whether a system lacks proactive communication capabilities or uses only limited tone and timing preferences. Additionally, FAICO can serve as a benchmarking tool, allowing comparisons between co-creative AI systems' AI communication. Evaluation tools and frameworks have been proposed and widely utilized in co-creativity domains \cite{karimi2018evaluating, rezwana2022designing}.

\subsection{Limitations \& Potential Improvements to Our Framework}
We acknowledge that our framework, in its current form, is in the preliminary stage, and there is room for refinement or extension. Additional aspects of communication and human factors essential for designing AI Communication in the co-creative context can be added to it. For example, in our framework, communication types include \emph{explanation}, \emph{suggestion}, and \emph{feedback}. However, insights from our focus groups prompted us to recognize emotional expression as a form of AI Communication as participants advocated for empathetic AI (Theme 6). More aspects of user experience, user demographics, or objective experience can be added to FAICO for future investigation.

Additionally, we do not claim generalizability in our findings of the preliminary study. However, we believe they suggest promising directions for future research. We also note potential biases in our participant sample as we recruit them from our personal networks and was relatively small although we strove to balance diverse expertise in the groups. Nonetheless, our framework usefully seems to represent a set of conceptually meaningful components of AI communication, is grounded by a systematic literature review, and prompts interesting insights from participants in our focus groups. Future work could also benefit from having non-expert perspective using larger user studies.

\section{Conclusion}

This paper introduces FAICO, a novel framework for designing AI communication that enhances user experience in human-AI co-creativity. Through a systematic literature review, we designed a framework that identifies the key aspects of AI Communication and their influence on user experience. A preliminary study to gather feedback on the framework identified insights such as the importance of context-awareness and flexible control over the AI's communication style, highlighting the need to accommodate diverse user preferences and different creative phases. Findings also show that a human-AI communication cycle is preferred over unidirectional communication for fostering mutual understanding between humans and AI, with participants wanting AI to communicate as an equal collaborator in the co-creation. These findings lay the foundation for designing human-centered co-creative AI systems and offer valuable directions for future research on improving AI communication to better support diverse users and foster meaningful collaboration.

\begin{acks}
Corey Ford was a student at Queen Mary University of London, supported by the EPSRC UKRI Centre for Doctoral Training in Artificial Intelligence and Music (AIM) [EP/S022694/1], at the inception of this work.
\end{acks}

\bibliographystyle{ACM-Reference-Format}
\bibliography{sample-base}

\appendix
\onecolumn
\section*{Appendix}
\vspace{-1em}

\begin{table}[H]
  \centering
  \caption{Focus group 1 participants' demographics and descriptions of their expertise. Note: P5 is removed as they couldn't contribute due to technical issues.}
  \resizebox{\textwidth}{!}{
    \begin{tabular}{lrlllp{33.5em}}
    \toprule
    \textbf{ID } & \multicolumn{1}{l}{\textbf{Age}} & \textbf{Gender} & \textbf{Job} & \textbf{Main Field} & \multicolumn{1}{l}{\textbf{Expertise \& Experiences}} \\
    \midrule
    P1    & 31    & M    & PhD Student & HCI  & Ph.D. in AI and Music with knowledge of generative AI. Background in performative arts, applying HCI expertise in music contexts. Design skills in participatory and soma design strategies. \\
    \midrule
    P2    & 27    & F   & PhD Student & HCI & In HCI, they specialize in conducting user studies and qualitative analysis. While not a design expert, they've done prototyping and UI/UX work. AI knowledge includes practical experience with AI models, e.g. Naive Bayes and LSTM. \\
    \midrule
    P3    & 35    & F   & Senior UX Researcher  & AI   & HCI researcher actively exploring generative AI tools with skills in implementing AI algorithms. For UX research, they investigate thoughts on generative AI tools among students and faculty. They have practical design experience designing prototypes. \\
    \midrule
    P4    & 32    & M   & Applied AI Scientist & AI    & 5+ years of experience in AI, Applied Machine Learning, NLP, and LLM. 2/3 years of experience in UX Research, survey design, and qualitative analysis. Experience forming design principles in PhD work. \\
    \midrule
    P6    & 22   & F      & Designer & Design  & Focused in UX design. Has a solid foundation with HCI modules from their degree program, primarily focusing on design and creative computing. Has experience in designing customer-facing chatbots from industry experience. \\
    \midrule
    P7    & 30    & F   & Student & HCI   & Moderate expertise in HCI research, possessing a foundational understanding of the subject. They are familiar with design thinking and processes. \\
    \bottomrule
    \end{tabular}%
  } 
  \label{tab:focusgroup1}%
\end{table}%

\begin{table}[H]
  \centering
  \caption{Focus group 2 participants' demographics and descriptions of their expertise.}
  \resizebox{\textwidth}{!}{
    \begin{tabular}{lrlllp{33.5em}}
    \toprule
    \textbf{ID } & \multicolumn{1}{l}{\textbf{Age}} & \textbf{Gender} & \textbf{Job} & \textbf{Main Field} & \multicolumn{1}{l}{\textbf{Expertise \& Experiences}} \\
    \midrule
    P8    & 39    & F    & University Faculty & AI    & Experience intersecting AI and Computer Science Education. Practical experience of Human-Centered Design through relevant courses taken during their PhD studies. \\
    \midrule
    P9    & 28    & F  & PhD researcher  & Design & 3.5 years of experience in HCI teaching and research. Studied Industrial Design for 5.5 years. Currently enrolled in a Design PhD program. \\
    \midrule
    P10   & 27    & M     & Student & HCI   & Intersection of AI and Machine learning. 3+ years of industry experience. Priovus experiences conducting diverse user studies on interaction design in the context of developer tools. \\
    \midrule
    P11   & 26    & F  & Student & HCI   & Theoretical knowledge of AI and hands-on experience of generative AI prompting. 4-5 years of HCI research experience, specializing in understanding human perception of AI and designing collaborative systems. Familiar with design principles. \\
    \midrule
    P12   & 34    & F    & Faculty & HCI   & AI-interaction researcher. Recent focus on a design study iteratively designed risk communication methods, applying HCI principles. \\
    \midrule
    P13   & 28    & M   & Student & AI    & This individual is an AI Ph.D. researcher with a Master's degree in HCI. They have extensive experience, having both undertaken multiple design courses and taught in the field. \\
    \bottomrule
    \end{tabular}%
    } 
  \label{tab:focusgroup2}%
\end{table}%

\end{document}